\documentclass[reprint,twocolumn,superscriptaddress,secnumarabic,amssymb,nobibnotes,aps,prx]{revtex4-2}

\usepackage{amsmath}    % need for subequations
\usepackage{graphicx}    % need for figures
\usepackage{verbatim}    % useful for program listings
\usepackage{color}           % use if color is used in text
\usepackage{subfigure}   % use for side-by-side figures
\usepackage{hyperref}    % use for hypertext links, including those to external documents 
\usepackage{verbatim}  % and URLs
\definecolor{ao}{rgb}{0.0, 0.5, 0.0}
\usepackage[normalem]{ulem}
\usepackage{lipsum}

\begin{document}
\title{Quantifying the U $5f$ covalence and degree of localization in U intermetallics}

\author{Andrea~Marino}
\affiliation{Max Planck Institute for Chemical Physics of Solids, N{\"o}thnitzer Stra{\ss}e 40, 01187 Dresden, Germany}
\affiliation{Institute of Solid State and Materials Physics, TU Dresden, 01069 Dresden, Germany}

\author{Denise~S.~Christovam}
\affiliation{Max Planck Institute for Chemical Physics of Solids, N{\"o}thnitzer Stra{\ss}e 40, 01187 Dresden, Germany}

\author{Daisuke~Takegami}
\affiliation{Max Planck Institute for Chemical Physics of Solids, N{\"o}thnitzer Stra{\ss}e 40, 01187 Dresden, Germany}
\affiliation{Department of Applied Physics, Waseda University, 3-4-1 Okubo, Shinjuku-ku, Tokyo 169-8555, Japan}

\author{Johannes~Falke}
\affiliation{Max Planck Institute for Chemical Physics of Solids, N{\"o}thnitzer Stra{\ss}e 40, 01187 Dresden, Germany}

\author{Miguel~M.~F.~Carvalho}
\affiliation{Max Planck Institute for Chemical Physics of Solids, N{\"o}thnitzer Stra{\ss}e 40, 01187 Dresden, Germany}
\affiliation{Institute of Physics II, University of Cologne, Z\"{u}lpicher Stra{\ss}e 77, 50937 Cologne, Germany}

\author{Takaki~Okauchi}
\affiliation{Department of Physics and Electronics, Osaka Metropolitan University 1-1 Gakuen-cho, Nakaku, Sakai, Osaka 599-8531, Japan}

\author{Chun-Fu~Chang}
\affiliation{Max Planck Institute for Chemical Physics of Solids, N{\"o}thnitzer Stra{\ss}e 40, 01187 Dresden, Germany}

\author{Simone~G.~Altendorf}
\affiliation{Max Planck Institute for Chemical Physics of Solids, N{\"o}thnitzer Stra{\ss}e 40, 01187 Dresden, Germany}

\author{Andrea~Amorese}
\altaffiliation{Present address: ASML Netherlands B.V., De Run 6501, 5504 DR, Veldhoven, The Netherlands.}
\affiliation{Max Planck Institute for Chemical Physics of Solids, N{\"o}thnitzer Stra{\ss}e 40, 01187 Dresden, Germany}
\affiliation{Institute of Physics II, University of Cologne, Z\"{u}lpicher Stra{\ss}e 77, 50937 Cologne, Germany}

\author{Martin~Sundermann}
\affiliation{Max Planck Institute for Chemical Physics of Solids, N{\"o}thnitzer Stra{\ss}e 40, 01187 Dresden, Germany}
\affiliation{PETRA III, Deutsches Elektronen-Synchrotron DESY, Notkestra{\ss}e 85, 22607 Hamburg, Germany}

\author{Andrei~Gloskovskii}
\affiliation{PETRA III, Deutsches Elektronen-Synchrotron DESY, Notkestra{\ss}e 85, 22607 Hamburg, Germany}

\author{Hlynur~Gretarsson}
\affiliation{PETRA III, Deutsches Elektronen-Synchrotron DESY, Notkestra{\ss}e 85, 22607 Hamburg, Germany}
\affiliation{Max Planck Institute for Solid State Research, Heisenbergstra{\ss}e 1, 70569 Stuttgart, Germany}

\author{Bernhard~Keimer}
\affiliation{Max Planck Institute for Solid State Research, Heisenbergstra{\ss}e 1, 70569 Stuttgart, Germany}

\author{Alexandr~V.~Andreev}
\affiliation{Institute of Physics, Academy of Sciences of the Czech Republic, Na Slovance 1999/2, 182 21 Prague 8, Czech Republic}

\author{Ladislav~Havela}
\affiliation{Department of Condensed Matter Physics, Faculty of Mathematics and Physics, Charles University, Ke Karlovu 5, 121 16 Prague 2, Czech Republic}

\author{Andreas~Leithe-Jasper}
\affiliation{Max Planck Institute for Chemical Physics of Solids, N{\"o}thnitzer Stra{\ss}e 40, 01187 Dresden, Germany}

\author{Andrea~Severing}
\affiliation{Institute of Physics II, University of Cologne, Z\"{u}lpicher Stra{\ss}e 77, 50937 Cologne, Germany}
\affiliation{Max Planck Institute for Chemical Physics of Solids, N{\"o}thnitzer Stra{\ss}e 40, 01187 Dresden, Germany}

\author{Jan~Kune{\v s}}
\affiliation{Department of Condensed Matter Physics, Faculty of Science, Masaryk University, Kotl\'a\v{r}sk\'a 2, 611 37 Brno, Czechia}

\author{Liu~Hao~Tjeng }
\altaffiliation{corresponding author: hao.tjeng@cpfs.mpg.de}
\affiliation{Max Planck Institute for Chemical Physics of Solids, N{\"o}thnitzer Stra{\ss}e 40, 01187 Dresden, Germany}

\author{Atsushi~Hariki}
\altaffiliation{corresponding author: hariki@pe.osakafu-u.ac.jp}
\affiliation{Department of Physics and Electronics, Osaka Metropolitan University 1-1 Gakuen-cho, Nakaku, Sakai, Osaka 599-8531, Japan}

\date{\today}

\begin{abstract}
	A procedure for quantifying the U $5f$ electronic covalency and degree of localization in U intermetallic compounds is presented. To this end, bulk sensitive hard and soft x-ray photoelectron spectroscopy were utilized in combination with density-functional theory (DFT) plus dynamical mean-field theory (DMFT) calculations. The energy dependence of the photoionization cross-sections allows the disentanglement of the U\,$5f$ contribution to the valence band from the various other atomic subshells so that the computational parameters in the DFT\,+\,DMFT can be reliably determined. Applying this method to UGa$_2$ and UB$_2$ as model compounds from opposite ends of the (de)localization range, we have achieved excellent simulations of the valence band and core-level spectra. The width in the distribution of atomic U\,$5f$ configurations contributing to the ground state, as obtained from the calculations, quantifies the correlated nature and degree of localization of the U\,5$f$. The findings permit answering the longstanding question why different spectroscopic techniques give seemingly different numbers for the U 5$f$ valence in intermetallic U compounds. 
\end{abstract}

\maketitle

\section{Introduction}
Intermetallic uranium compounds display a plethora of intriguing phenomena including heavy-fermion behavior, unconventional and spin-triplet superconductivity, hidden and multipolar ordering, singlet magnetism, as well as coexistence of ordered phases~\cite{pfleiderer2009,brando2016,Mydosh2020,Lewin2023}. It is generally accepted that these phenomena originate from the competition and interplay of U\,$5f$ local atomic-like correlation effects and the tendency of the U $5f$ to delocalize and form bands. The U $5f$ covalence and degree of localization are hereby key indicators that characterize the system. Yet, the assessment of the electronic structure of these uranium intermetallic compounds is truly challenging since one must go beyond mean-field approaches.

Recent developments based on density functional theory (DFT) combined with dynamical mean field theory (DMFT) provide an exciting window of opportunity to capture certain aspects of the low and high energy properties for a number of U compounds~\cite{haule2009,kung2015,miao2019,miao2020}. Although achieving very promising results, DFT\,+\,DMFT calculations often face the problem of, for example, how to subtract correlation effects already included in both the DFT and the DMFT parts~\cite{Haule2015} and how to determine the actual magnitude of the correlations. Consequently, accurately reproducing experimentally observed quantities, such as spectra in electron spectroscopy, may pose difficulties if the calculations were to be truly \textit{ab initio}. As a result, the community is facing seemingly conflicting results from different spectroscopic techniques regarding the extent of correlations and the occupation of the 5$f$ shell. Examples include the well-known hidden order and unconventional superconductor URu$_2$Si$_2$\,\cite{Jeffries2010,Fujimori2012,Wray2015,Booth2016,Sundermann2016,Kvashnina2017}, and the recently discovered  spin-triplet superconductor UTe$_2$\,\cite{Fujimori2019UTe2,miao2020,Thomas2020,fujimori2021,Shick2021,Aoki2022b,liu2022,Wilhem2023,Christovam2024}.

\begin{figure*}[t!]
	\includegraphics[width=0.9\linewidth]{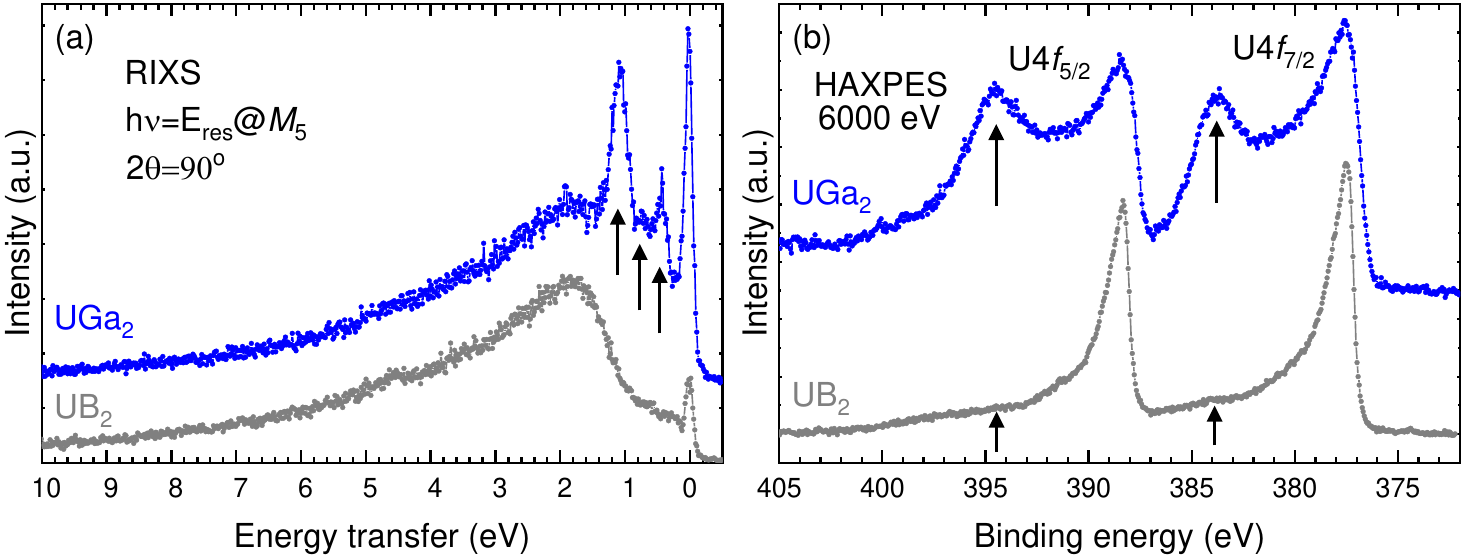}
	\caption{(a) U $M_5$-edge RIXS spectra of UGa$_2$ as adapted from Ref.~\cite{marino2023} and of UB$_2$, both measured at resonance energy. The arrows indicate multiplet excitations. (b) HAXPES U 4$f$ core level spectra after Shirley background subtraction of UGa$_2$ and UB$_2$ measured with 6000\,eV incident energy, with arrows pointing out the satellite structures. }
	\label{RIXS_HAXPES}
\end{figure*}

 Here we aim to establish a procedure to empirically reduce or even effectively eliminate the uncertainties inherent in DFT\,+\,DMFT. Material specific parameters, such as the so-called double counting correction $\mu_{\rm dc}$, the Hubbard $U_{\rm ff}$ and Hund's $J$ of the U\,$5f$ states, are obtained by tuning the simulation to accurately reproduce the valence band (VB) spectra that are measured with photoelectron spectroscopy (PES) using different photon energies. Measuring the VB with different photon energies allows enhancing or suppressing contributions of the various subshells because their photoionization cross-sections depend differently on the incident energy~\cite{TRZHASKOVSKAYA2001,TRZHASKOVSKAYA2002,TRZHASKOVSKAYA2006}; the U $5f$ states contribute most strongly when the VB is measured with soft x-rays whereas they are strongly suppressed in hard x-ray spectra. Having disentangled the spectral weights of the subshells, the parameters in the DFT\,+\,DMFT calculation can be optimized.

We consider isostructural UGa$_2$ and UB$_2$ as model materials that are at the opposite sides of the localization spectrum. UGa$_2$ is expected to be more atomic-like, while UB$_2$ must be more delocalized, as inferred from the U--U distances $d_\text{UU}$ in the their simple AlB$_2$ hexagonal structure (space group P6/\textit{mmm})~\cite{makarov1956,andreev1978, dancausse1992}. With $d_\text{UU}$\,=\,4.212\,\AA\,  UGa$_2$ is well above the Hill limit of 3.5\AA\ \cite{hill1970}, whereas UB$_2$ with $d_\text{UU}$\,=\,3.123\,\AA\  is well below. The observed high-moment ferromagnetism (3\,$\mu_B$)~\cite{lawson1985, honma2000} in UGa$_2$ and the paramagnetic properties of UB$_2$ down to the lowest measured temperature~\cite{yamamoto1999} support this assumption. Furthermore, the VB band resonant inelastic x-ray scattering (RIXS) spectra of UGa$_2$, measured at the U $M_5$ edge, exhibit sharp multiplet excitations (see black arrows in Fig.\,\ref{RIXS_HAXPES}\,(a)). These excitations were analyzed using a full multiplet approach, which showed that the main configuration is $5f^2$ with $\Gamma_1$ and/or $\Gamma_6$ symmetry. It was further shown, that the large in-plane magnetic moment of UGa$_2$ is of the induced type, with the $\Gamma_1$ singlet being lowest in energy and coupled to the higher-lying $\Gamma_6$ doublet\,\cite{marino2023}.  In contrast, the RIXS spectra of UB$_2$ only show charge transfer scattering, whereas the multiplet excitations are seemingly so strongly broadened that they are no longer distinguishable. Moreover, the U 4$f$ core level spectra obtained with hard x-ray photoelectron spectroscopy (HAXPES) exhibit stark differences; in UGa$_2$, the main emission lines at 378 and 390eV binding energy show prominent satellites, while they are barely visible in the UB$_2$ core-level data (see black arrows in Fig.\,\ref{RIXS_HAXPES}\,(b)). The present core-level data agree well with soft x-ray investigations (see \,\cite{schneider1981, gouder2001, fujimori2019} for UGa$_2$ and\,\cite{fujimori2016} for UB$_2$). For experimental details of the RIXS and HAXPES data, we refer to the section II\,Methods and for further discussion of the spectra we refer to sections III\,Results and IV\,Discussion. 

This work aims to provide a quantitative analysis of the charge state of uranium intermetallic compounds. By employing a cross-section guided VB PES study combined with  DFT\,+\,DMFT calculations, we will determine the charge distribution, the average valence and the degree of covalence for UGa$_2$ and UB$_2$. Using these results, we will \textit{quantitatively} compute the U\,4$f$ core-level spectra. Additionally, we will discuss the presence and absence of multiplet states in the UGa$_2$ and UB$_2$ $M_5$-edge RIXS spectra, respectively, and address the discrepancies in the interpretation of different spectroscopy techniques concerning the valence states in this material class.

%*******************************************
%*******************************************
%*******************************************
\section{Methods}
%*******************************************
%*******************************************
%*******************************************
\subsection{Experiment}
UGa$_2$ single crystals were grown with the Czochralski method and characterized prior to the experiment\,\cite{kolomiets2015}. Polycrystalline samples of UB$_2$ were synthesized by arc-melting of stoichiometric amounts of crystalline boron (Chempur, 99.95\%) together with natural uranium metal (Goodfellow, 99.9\%) under a protective atmosphere of Ar-gas on a water cooled copper-hearth. The arc- melted sample was placed in a alumina crucible, welded into a tantalum tube and annealed at 1100$^o$\,C for 48 hours.

HAXPES measurements were conducted at the P22 beamline\,\cite{schlueter2019} at PETRA III (DESY) in Hamburg, Germany. The incident photon energy was set to 6000\,eV, thus assuring a higher bulk sensitivity, and photons were detected with a SPECS225HV electron analyzer in the horizontal plane at 90$^{\circ}$ from the incoming beam, with the sample surface 45$^{\circ}$ from the incoming beam.  Photoelectron spectroscopy with soft x-rays (SXPES) was measured with 1200\,eV photon energy at the NSRRC-MPI TPS 45A Submicron Soft X-ray Spectroscopy beamline\,\cite{huangming2019} at the Taiwan Photon Source. Here, the photoelectrons were collected by an MB Scientific A-1 photoelectron analyzer in the horizontal plane at 60$^{\circ}$ with sample emission along the surface normal. SXPES with 1486.7\,eV photon energy were performed with monochromatized Al\,K$\alpha$ light in normal emission geometry. Electrons were collected with a Scienta R3000 electron analyzer. In all cases, clean sample surfaces were achieved by cleaving \textit{in situ} in ultra-high-vacuum condition. The pressures of the main chambers were in the low 10$^{-10}$\,mbar range. Valence band spectra of a gold or silver sample were used for calibrating the Fermi level and the overall energy resolution, which was $\approx230$\,meV for the for the 6000\,eV, $\approx200$\,meV for the 1200\,eV  and $\approx300$\,meV for the 1486.7\,eV data. The sample temperature $T$ was kept at 40~K for the 6000\,eV and 1200\,eV, and at 77\,K for the 1486.6\,eV measurements. 

The RIXS experiments at the U\,$M_5$-edge were performed at Max-Planck IRIXS endstation of P01 beamline at Petra III/DESY (Hamburg). The RIXS instrument utilizes hard x-ray optics, as described in Ref.\,\cite{gretarsson2020}, and achieves a resolution of 150\,meV at the U\,$M_5$-edge with the 112 reflection of a diced quartz analyzer, as estimated by measuring a carbon tape. The experiment was performed with a scattering angle $2\theta=90^o$ to minimize elastic scattering, a sample angle $\theta_s=45^o$, and $T$\,=\,35\,K. 

\subsection{Calculations}
The valence-band and core-level PES simulations started with a standard DFT\,+\,DMFT calculation~\cite{kotliar06,georges96} with the local density approximation (LDA) for the exchange correlation potential.  First, the DFT bands for the experimental crystal structure were obtained by the Wien2K package~\cite{wien2k} and subsequently projected onto a tight-binding model spanning U\,5$f$, 7$s$, 6$d$ orbitals, Ga\,4$s$, 4$p$ and B\,2$s$, 2$p$ orbitals using the wien2wannier and wannier90 packages~\cite{wien2wannier,wannier90}. The spin-orbit coupling (SOC) was included in the DFT calculation. The tight-binding model was augmented with a local electron-electron interaction within the U\,5$f$ shell, parameterized by the Hubbard $U_{\rm ff}$ and Hund’s $J$ parameters. 
The strong-coupling continuous-time quantum Monte Carlo (CT-QMC) impurity solver~\cite{werner06,gull11,boehnke11,hafermann12} was used to compute the U\,5$f$ self-energy, which was analytically continued to real frequency after the convergence of the DMFT self-consistent loop was achieved~\cite{jarrell96}. For computational efficiency, only the density-density terms in the Coulomb interaction vertex were taken into account in the CT-QMC calculation. The calculations were performed for $T=300$~K.

Merging the DFT scheme with DMFT faces the well-known problem of how to avoid double-counting the U 5$f$--5$f$ interaction already present in the DFT results~\cite{kotliar06,Karolak10,Haule15}. To this end, one needs to determine the U\,5$f$ site energies by subtracting the so-called double-counting correction $\mu_{\rm dc}$ from the respective DFT values, but a generally accepted universal expression for $\mu_{\rm dc}$ is not available. Practically, $\mu_{\rm dc}$ renormalizes the energy splittings between U\,5$f$ and other uncorrelated orbitals. 

By successively varying $\mu_{\rm dc}$ in the DFT\,+\,DMFT calculations for certain sets of $U_{\rm ff}$ and $J$, the best agreement of prominent valence-band features in the experimental valence band PES data with the DFT\,+\,DMFT spectra determines the $\mu_{\rm dc}$ value for the correction. The double-counting dependence of the valence-band spectra is shown in detail for different sets of $U_{\rm ff}$ and $J$ in Appendix~\ref{appendix_U_and_dc_scans}.

\begin{figure*}[t!]
	\center
	\includegraphics[width=1\linewidth]{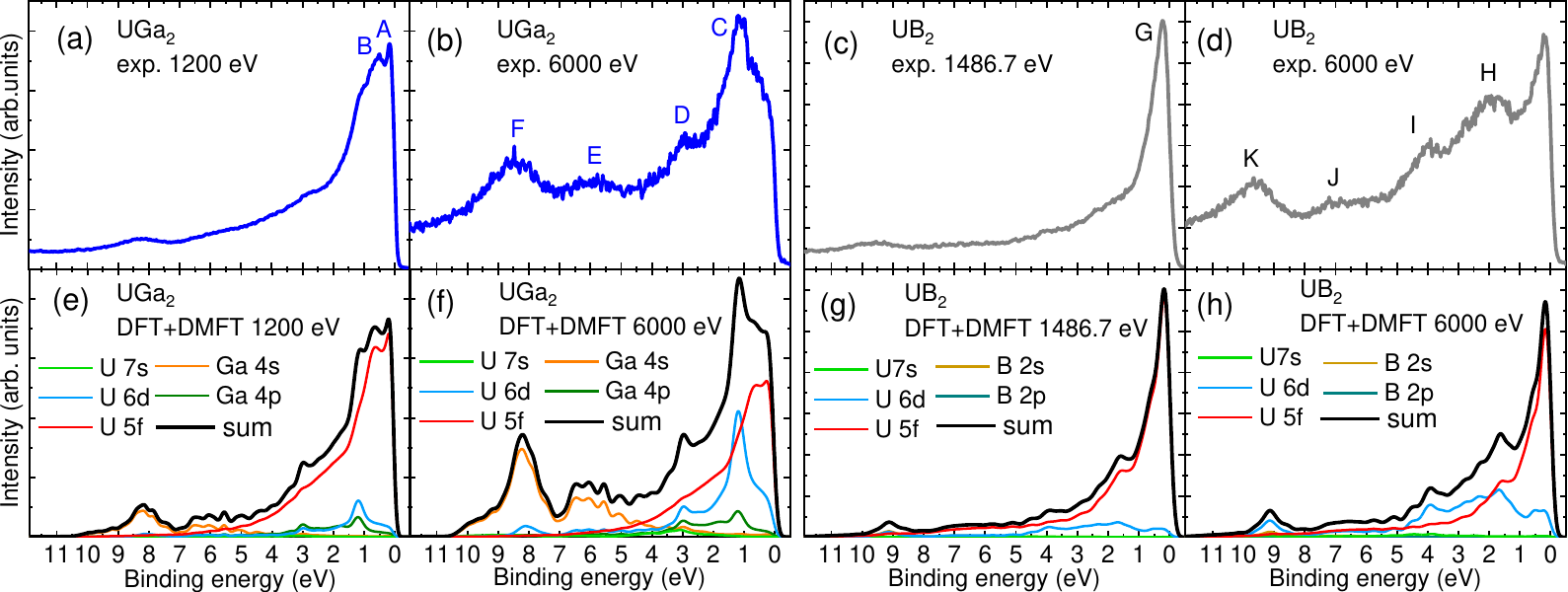}
	\caption{(a)-(d) Experimental hard and soft x-ray valence band spectra of UGa$_2$ and UB$_2$ measured with 1200\,eV, 1486.7\,eV and 6000\,eV photon energies. (e)-(h) DFT+DMFT calculations of the valence band spectra of UGa$_2$ and UB$_2$ for $U_{\rm ff}=3$\,eV, $J=0.59$\,eV and $\mu_{\rm dc}$\,=\,4.25\,eV, after photoionization cross-sections corrections (see text) and Gaussian broadening. }
	\label{fig_VB}
\end{figure*}

\begin{figure}[]
	\includegraphics[width=0.99\columnwidth]{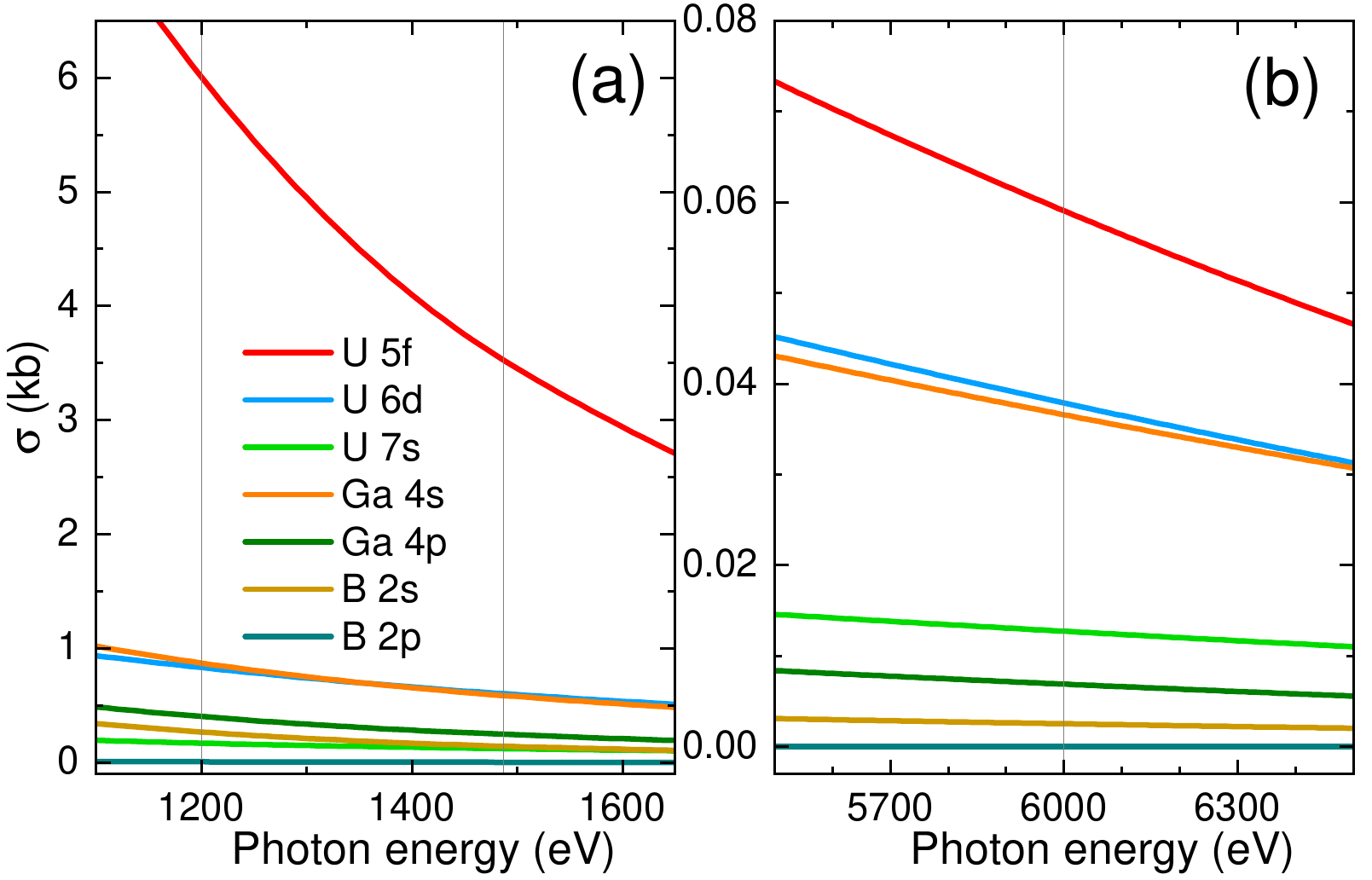}
	\caption{Photoionization cross sections of U\,5$f$, U\,6$d$, U\,7$s$, Ga\,4$s$, Ga\,4$p$, B\,2$s$, B\,2$p$ as interpolated from Refs.\,\cite{TRZHASKOVSKAYA2001, TRZHASKOVSKAYA2002, TRZHASKOVSKAYA2006}. The vertical lines indicate the photon energies where the experiments were performed.}
	\label{fig_sigma}
\end{figure}

The core-level PES spectra were calculated from the Anderson impurity model (AIM) augmented with the core states and the real-frequency DFT\,+\,DMFT hybridization function discretized into 23~levels (per spin and orbital). For this purpose, we employed the configuration-interaction impurity solver, which allowed us to treat a large number of bath levels (see Refs.\,\cite{Hariki17,Winder20} for details) compared with a standard exact diagonalization solver~(see, e.g., Ref.\,\cite{Chatterjee21}). Nevertheless, treating full U\,4$f$ core-orbital degrees of freedom was a computationally demanding task. Thus, $s$-type ($l=0$) core orbitals were adopted in the AIM, meaning the SOC splitting on the U\,4$f$ core level and the multipole part in the core-valence (U\,$4f$--$5f$) Coulomb interaction were neglected in the spectra. The U\,4$f_{5/2}$ and 4$f_{7/2}$ SOC splitting is large ($\sim 10$~eV), and consequently interference effects between the two SOC partners are negligibly small. Thus, only the monopole term $U_{\rm fc}$ in the core-valence interaction had to be adjusted for reproducing the experimental spectra.

%*******************************************
%*******************************************
%*******************************************
\section{Results}
%*******************************************
%*******************************************
%*******************************************
Figure~\ref{fig_VB}\,(a) to (d) shows the experimental valence-band PES spectra of UGa$_2$ and UB$_2$ measured with soft and hard x-rays. Our low photon energy data agree well with previously reported soft x-ray studies\,\cite{fujimori2019, fujimori2016}. We identify features labeled as $A$ to $K$ in the spectra. These features do not shift in energy but exhibit different spectral weights depending on the incident photon energy. In UGa$_2$, the double peak structure, $A$ (0.15\,eV) and $B$ (0.5\,eV), is pronounced in the 1200~eV data, while features $C$ (1.11\,eV) to $F$ are prominent in the 6000~eV spectrum. It is important to note that feature $C$ has a higher binding energy than feature $B$. Likewise in UB$_2$, feature $G$ emerges in the 1486.7~eV data, while the spectral weights of features H to K are enhanced in the 6000~eV dataset. 

The energy dependence of the photoionization cross-sections\,\cite{YEH1985, TRZHASKOVSKAYA2001, TRZHASKOVSKAYA2002, TRZHASKOVSKAYA2006} can be exploited to disentangle the different orbital contributions in the valence-band spectra. Figure~\ref{fig_sigma}\,(a) and (b) shows the photoionization cross sections of the Ga\,4$s$, Ga\,4$p$, B\,2$s$, B\,2$p$, U\,7$s$, U\,6$d$, and  U\,5$f$ orbitals in the soft and hard x-ray energy range respectively, with the vertical lines signaling the photon energies used in the experiments.  The cross-sections generally decrease with increasing photon energy, with the U\,5$f$ one always being the largest. However, in the soft x-ray range, the ratio of the U\,5$f$ spectral weights to the weights of the other states is at least 3.8 times larger than at 6000~eV. We can thus conclude that the features enhanced in the soft x-ray data, $A$, $B$ in the UGa$_2$ spectrum, and $G$ in the UB$_2$ spectrum, are contributions from U\,5$f$, while all the other features are due to contributions from non-5$f$ states. 

Figure~\ref{fig_VB}(e) to (h) shows the optimized DFT\,+\,DMFT calculation for the valence band spectra of UGa$_2$ and UB$_2$, after weighing the individual orbital intensities with the respective cross-sections in Fig.\,\ref{fig_sigma} and after some broadening to mimic the respective experimental resolutions. The parameter values of $U_{\rm ff}=3$\,eV, $J=0.59$\,eV and $\mu_{\rm dc}$\,=\,4.25\,eV reproduce best $all$ the characteristic valence-band features ($A$--$K$) at two photon energies simultaneously (compare to Fig.~\ref{fig_VB}(a) to (d)). In Appendix~\ref{appendix_U_and_dc_scans}, we describe the parameter search and demonstrate how the $\mu_{\rm dc}$ parameter scales almost linearly with the relative position of the features originating from the U\,5$f$ states and the features $F$ and $K$ that come from uncorrelated states. We pick the value of $\mu_{\rm dc}$ that fits best the energy position of features $F$ and $K$, as well as the line shape of the correlated U\,$5f$ spectrum. The presence of the double-peak features $A$ and $B$ in the U\,5$f$ states is a manifest of the sizable Hund $J$ interaction. 

The comparison of valence spectra and simulations allows identifying the dominant orbital contributions:~features $A$, $B$ in panel (a) and $G$ in panel (c) indeed mainly originate from the U\,$5f$ states. The intensities $C$ and $D$ in the spectra of UGa$_2$ mainly come from the U $6d$ and intensities E and F from the Ga\,$4s$ orbitals. For UB$_2$ features $H$ and $I$ are primarily due to the U\,$6d$, while features $J$ and $K$ also have a sizable B\,$2s$ contribution.

%%%%%%%%%%%%%%%%%%%%%%%%%%%%%%%%%%%%%%%%%%%

 \begin{figure}[t!]
\includegraphics[width=0.75\columnwidth]{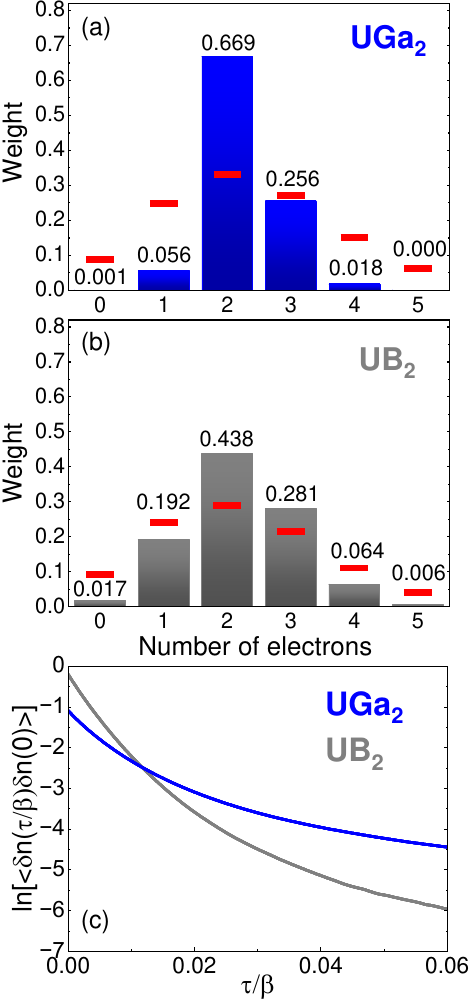}
    \caption{The atomic configuration histograms of the U\,5$f$ states in UGa$_2$ (a) and UB$_2$ (b) calculated by the DFT\,+\,DMFT method with optimized parameters, $U_{\rm ff}$\,=\,3\,eV, $J$\,=\,0.59\,eV and $\mu_{\rm dc}$\,=\,4.25\,eV. The red ticks indicate an uncorrelated statistical distribution of atomic configurations (see text). (c) Time dependent charge correlation functions $\langle \delta n(\tau/\beta) \delta n(0)\rangle$ of UGa$_2$ and UB$_2$ calculated by the DFT\,+\,DMFT method. }
\label{Fig_histograms_and_charge_corr}
\end{figure}

Given the good agreement of the DFT\,+\,DMFT calculations with the valence-band experimental spectra, we now assess the U valence. To this end, we resort to a valence histogram on the uranium site, using the DFT\,+\,DMFT method with the optimized material specific parameters. These histograms represent projections of the local density matrix on different 5$f$ occupation number sectors. It is computed from the AIM using the numerically-exact CT-QMC solver with the DMFT hybridization densities $\Delta (\omega)$. In Figure~\ref{Fig_histograms_and_charge_corr}~(a) and (b), we present the resulting histograms of UGa$_2$ and UB$_2$. It is evident that numerous 5$f^n$ atomic configurations contribute to the respective ground states of UGa$_2$ and UB$_2$. Thus, the U nominal valence or dominant (starting) configuration for an ionic picture is unclear and is not known \textit{a priori} for U intermetallics. The $\mu_{\rm dc}$ shifts the (bare) U\,5$f$ energy, and thus has a strong effect on the U valence distribution. A smaller $\mu_{\rm dc}$ results in a shallower U $5f$ level, thereby leading to smaller fillings. We show the variation of the histograms as a function of $\mu_{\rm dc}$ in Appendix~\ref{appendix_histograms}.

In both the histograms of UGa$_2$ and UB$_2$, the U\,5$f^2$ and 5$f^3$ configurations are the most prominent. In UGa$_2$, the $5f^1$, $f^2$, $f^3$ and $f^4$ configurations account for 6\,\%, 66\,\%, 26\,\% and 2\,\%  respectively, resulting in a filling of the U\,$5f$ shell of $\langle n \rangle =n_{5f}=2.24$. In UB$_2$, the $f^0$ and $f^5$ configurations also contribute. We find 1.7\% of $f^0$, 19.2\% of $f^1$, 43.8\% of $f^2$, 28.1\% of $f^3$, 6.4\% of $f^4$ and 0.6\% of $f^5$, resulting in a filling of $\langle n \rangle$\,=\,$n_{5f}$\,$\approx$\,2.20. Despite the very different physical properties of  UGa$_2$ and UB$_2$, their \textit{average} $5f$ shell fillings are remarkably similar.

\begin{figure}[]
	\includegraphics[width=0.8\columnwidth]{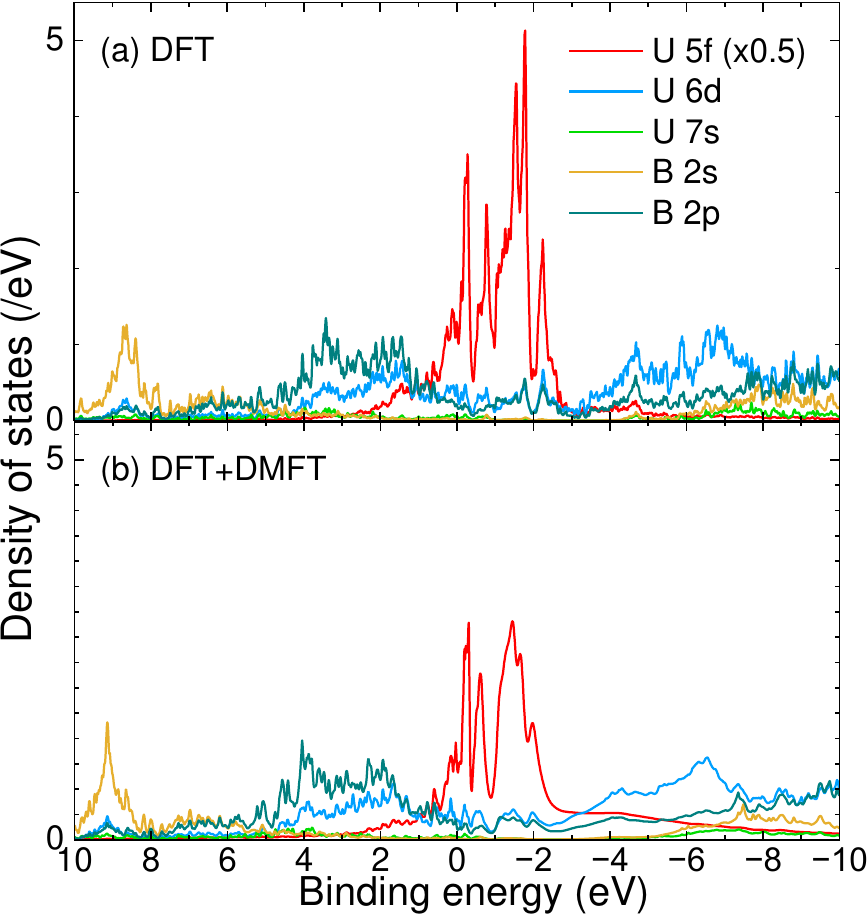}
	\caption{(a) DFT and (b) DFT\,+\,DMFT (with $\mu_{\rm dc}$\,=\,4.25\,eV) density of states of UB$_2$.}
	\label{Fig_DFT_DFTdmft_ub2}
\end{figure}

The key distinction between the histograms in Figure~\ref{Fig_histograms_and_charge_corr}~(a) and (b) lies in the width of the distributions. In UGa$_2$, the histogram peaks very narrowly around the $f^2$ configuration, while in UB$_2$, the distribution is much broader. The width of the histogram carries information about the magnitude of the charge fluctuations, which we find to be significantly larger in UB$_2$. We also add the statistical (binomial) distribution of the charge configurations that correspond to a completely itinerant uncorrelated $f$ shell (see red ticks in Figure~\ref{Fig_histograms_and_charge_corr}~(a) and (b)). The binomial distribution is given by 
\begin{equation}
	P(f^n)=\binom{14}{n}(1-c)^{14-n}c^n
\end{equation}
where $c$ is the electron concentration $c=n_f/14$.~\cite{phdthesis_bosch}. It shows that atomic correlations are not entirely absent in UB$_2$ because its DFT\,+\,DMFT histogram of UB$_2$ is still not as broad as the binomial distribution. Nevertheless, the distribution is wide enough so that the DFT\,+\,DMFT results yield only small spectral changes in comparison to the non-interacting DFT results, as shown in Fig.\,\ref{Fig_DFT_DFTdmft_ub2}.

Above, we inferred the average $5f$ occupation $n_{5f}$ as well as the magnitude of charge fluctuations $\langle \delta n \delta n\rangle$ directly from the histograms. However, charge fluctuations can have rather different origins such as itinerancy, i.e., formation of (broad) bands in intermediate valence systems or the degeneracy of valence states -- mixed valence. In order to distinguish these situations time dependent charge correlation functions $\langle \delta n(\tau) \delta n(\tau')\rangle$ must be evaluated~\cite{Ylvisaker2009}. In the former case, the correlation function rapidly decays with time $\tau-\tau'$, while in the latter the correlation function contains a long-lived component (in the limiting case of valence degeneracy of isolated atoms, the correlation function remains constant over time). In Figure~\ref{Fig_histograms_and_charge_corr}~(c), we show the local charge-charge correlation function $\langle \delta n(\tau) \delta n(0)\rangle$ in the imaginary time domain for UGa$_2$ and UB$_2$. We notice that the value of the function at $\tau=0$ is larger for UB$_2$ than for UGa$_2$, indicating larger charge fluctuations in the former compound. Moreover, the function decays faster in UB$_2$, suggesting that the charge fluctuations in this compound are primarily due to itinerancy. For a material even more itinerant than UB$_2$, the charge correlation function would decay even more rapidly over time (and the histogram resembling more the binomial distribution), whereas for a material even more localized than UGa$_2$, it would eventually become a constant over time (and the histogram becoming even narrower). Such a comparison would be particularly useful for the relative classification of several other intermetallics.

\begin{figure}[t!]
	\includegraphics[width=1\columnwidth]{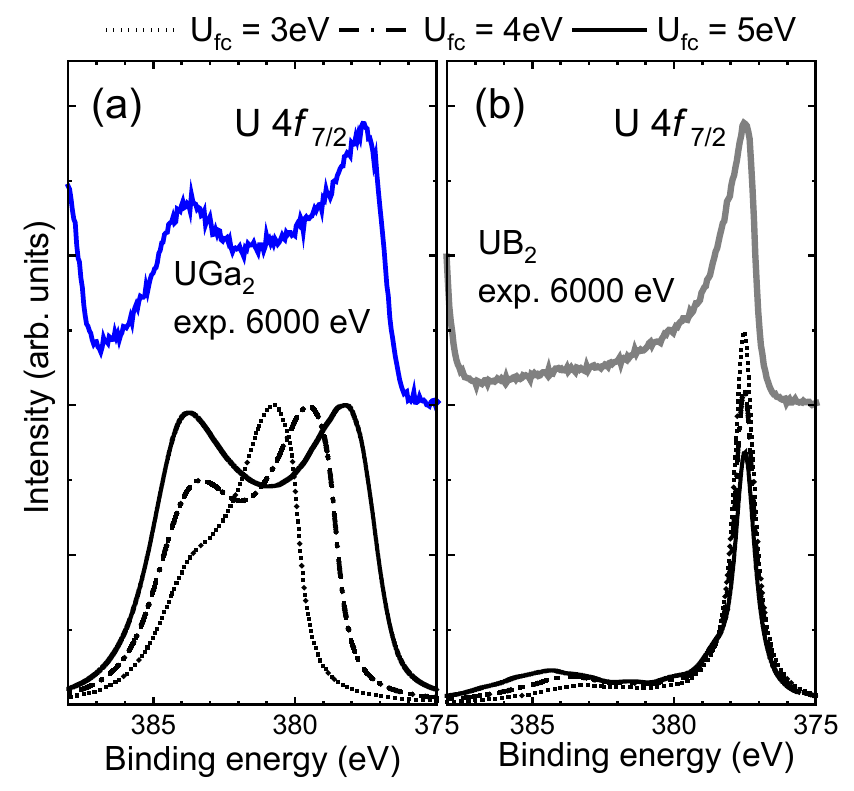}
\caption{U\,4$f_{7/2}$ core-level HAXPES spectra of UGa$_2$ (a) and UB$_2$ (b), compared to calculations performed with the DFT\,+\,DMFT AIM for different core-valence interaction values $U_{\rm fc}$. }
	\label{Fig_core_level}
\end{figure}

We now turn our attention to the U\,4$f$ core-level excitations of UGa$_2$ and UB$_2$, measured with 6000\,eV photon energy, after subtraction of a Shirley-type integral background (see in Fig.\,\ref{RIXS_HAXPES}\,(b)). Figure~\ref{Fig_core_level}~(a) and (b) display the U\,4$f_{7/2}$ core-level spectra on an enlarged scale. In the case of UGa$_2$, a prominent satellite peak is observed at $\approx6$\,eV above the main peak. It is important to note that this feature is not present in the U\,$5f$ spectral weight in the valence-band spectra, see Fig.~\ref{fig_VB}~(a). In contrast, the spectrum for UB$_2$ closely resembles that of an (almost) uncorrelated metal: a main, asymmetric peak is observed with minimal to no satellite structure. We then computed the U\,4$f$ core-level spectra using the DFT\,+\,DMFT AIM~\cite{Hariki17,Winder20}. The AIM incorporates the DFT\,+\,DMFT hybridization densities $\Delta(\omega)$ for the valence-band structure determined above. The sole tuning parameter in this computation is the amplitude of the interaction $U_{\rm fc}$ between the highly-localized U\,4$f$ core hole generated by the x-rays and the $5f$ valence electrons. Figure~\ref{Fig_core_level}~(a) and (b) show the calculated spectra for both compounds for three values of $U_{\rm fc}$. 

An excellent reproduction of the energy positions as well as ratios of main and satellite peaks is achieved with $U_{\rm fc}$\,=\,5\,eV, 
which, for the first time, defines this parameters in uranium core-level spectroscopy. It turns out that the core level data of both compounds are well described with the same set of parameters $U_{\rm fc}$, $U_{\rm ff}$, $\mu_{\rm dc}$ and $J$. Hence, the difference in the spectra comes from the larger itinerancy in U$B_2$, as revealed by the DFT\,+\,DMFT calculation. In Appendix\,\ref{appendix_toymodel} we provide a demonstration of the impact of itinerancy on the core-level spectra.

\section{Discussion}

 The accurate description of the VB as well as U\,4$f$ core-level spectra lends credibility to the choice of parameters, especially the double counting correction $\mu_{\rm dc}$, in the present DFT\,+\,DMFT calculation. Our results show that the U\,5$f$ electrons in UB$_2$ are much more itinerant than in UGa$_2$. Nevertheless, U\,5$f^2$ configuration is the most abundant the average valence ($n_{5f}$\,$\approx$\,2.2) is almost identical in both compounds. The latter fact shows how treacherous the qualitative interpretation of U\,4$f$ core-level data in terms of valence can be (see Appendix\,\ref{appendix_toymodel}). The $M_5$-edge RIXS, on the other hand, do reflect the stronger localization of the 5$f$ states in UGa$_2$ in a very obvious manner: the UGa$_2$ RIXS data exhibit multiplet excitations, whereas these are washed out for UB$_2$ due to the significantly larger itinerancy of its 5$f$ electrons.

The present results differ from a DFT\,+\,DMFT study of UGa$_2$ by Chatterjee \textit{et al.}\,\cite{Chatterjee21}. Aiming to explain the magnetism, they chose the double-counting parameter $\mu_{\rm dc}$ so that the 5$f$ filling in the DFT\,+\,DMFT result is equal to its DFT value. While Chatterjee \textit{et al.} also find a double peak structure close to the Fermi energy, the assignment of features $B$ and $C$, originating from U\,5$f$ and other states, respectively, is interchanged (see Fig.\,\ref{fig_VB}\,(a)). In our work the assignment is supported by the strong contrast between SXPES and HAXPES, the latter enhancing the non-5$f$ spectral weights as shown in Fig.\,\ref{fig_VB}\,(b).

Next we will address the issue of conflicting results from different spectroscopic techniques for U intermetallic compounds.
Having established the distinct presence of the electronic configurations involved, as well as having determined the magnitude
of the electronic correlation energies $U_{\rm fc}$ and $U_{\rm ff}$, we now are in the position to  use the language of
configuration interaction to resolve this issue, thereby highlighting a central aspect in the DMFT approach. The histograms presented in Figure~\ref{Fig_histograms_and_charge_corr} show that multiple $5f$ valence configurations contribute to the ground state, with  $5f^2$ and $f^3$ carrying the largest weight. Therefore, we focus on these two configurations.  How does the U\,5$f$ mixed valence state with $5f^2$ and $f^3$ configurations, e.g. as in UGa$_2$, manifest itself in different spectroscopic techniques and why do different techniques seemingly give different answers?  

In the following, we consider two typical core-level spectroscopy cases~\cite{gunnarsson1983, Kotani1988, degroot1994}:~1) In processes conserving charge neutrality, such as x-ray absorption (XAS) or non-resonant inelastic x-ray scattering (NIXS) at the $M_{4,5}$, $N_{4,5}$, $O_{4,5}$ edges of uranium, the energy ordering of the states originating from the $5f^2$ and $f^3$ configurations is the same with (final state) and without (initial state) the core hole. The key parameter for the energy order of the final states is the core-valence interaction $U_{\rm fc}$, which is comparable to the valence Coulomb interaction $U_{\rm ff}$. In these charge-neutral processes, where the core-electron excitation takes place locally at the x-ray irradiated site resulting in formation of an exciton, the energy gain attributed to the $U_{\rm fc}$ term is primarily offset by energy loss from the $U_{\rm ff}$. Thus the perturbation presented by the x-ray irradiation renormalizes only slightly the energy difference of the underlying ionic configurations. In this case, the spectral weights are \textit{biased} towards the dominant configuration thus it should not come as much of a surprise that U\,$O_{4,5}$ NIXS and $M_{5}$ RIXS spectra of UGa$_2$ have recently been well simulated with an ionic U\,5$f^2$ model approach\,\cite{marino2023}. We may say this kind of methods are mostly sensitive to the main configuration and relevant symmetries of the system. We note, that also an atomic 5$f^2$ configuration can explain the magnetism in UGa$_2$\,\cite{richter1997,honma2000,marino2023}.   

2) Electron removal processes, like PES, promote an inversed order  of the ionic configurations in the final state, due to the uncompensated core-hole potential $U_{\rm fc}$. These spectroscopies are characterized by shake-up and satellite structures. In case of U intermetallics, the spectral weights are determined by both the 5$f^2$ and 5$f^3$ configurations, making such a spectroscopy truly sensitive to covalence. 

As a service to the reader, we provide a simple demonstration of these effects on the spectra in Appendix~\ref{appendix_toymodel}. We construct a simple two-level toy-model for UGa$_2$, and present the effects of the relevant intra- ($U_{\rm ff}$) and inter-shell ($U_{\rm fc}$) Coulomb interactions on the the final state total energy diagram and on the redistribution of spectral weights. We also consider the effect that an increased hopping has on the toy-model, so as to mimic a more itinerant system like UB$_2$. We conclude, the hybridization and covalence effects in U intermetallics are so strong that simple cluster calculations can no longer describe the PES core-level spectra as, e.g., for Ce intermetallic compounds (see, e.g., Refs.~\cite{gunnarsson1983,Kotani1988,degroot1994,Sundermann2016a,Shimura2021}). For U intermetallics, the analysis of the PES core-level spectra is no longer intuitive and, as the present work has shown, requires a full DFT\,+\,DMFT based AIM calculation.

%%%%%%%%%%%%%%%%%%%%%
%%%%%%%%%%%%%%%%%%%%%%%%%%%%%%%%%%%%%%%%%%%%%%%%%%%%%%%%%%%%%
%%%%%%%%%%%%%%%%%%%%%%%%%%%%%%%%%%%%%%%%%%%%%%%%%%%%%%%%%%%%%

%%%%%%%%%%%%%%%%%%%%%%%%%%%%%%%%%%%%%%%%%%%%%%%%%%%%%%%%%%%%%
%%%%%%%%%%%%%%%%%%%%%%%%%%%%%%%%%%%%%%%%%%%%%%%%%%%%%%%%%%%%%
%%%%%%%%%%%%%%%%%%%%%%%%%%%%%%%%%%%%%%%%%%%%%%%%%%%%%%%%%%%%%

%%%%%%%%%%%%%%%%%%%%%%%%%%%%%%%%%%%%%%%%%%%%%%%%%%%%%%%%%%%%%
%%%%%%%%%%%%%%%%%%%%%%%%%%%%%%%%%%%%%%%%%%%%%%%%%%%%%%%%%%%%%
%%%%%%%%%%%%%%%%%%%%%%%%%%%%%%%%%%%%%%%%%%%%%%%%%%%%%%%%%%%%%

%%%%%%%%%%%%%%%%%%%%%%%%%%%%%%%%%%%%%%%%%%%%%%%%%%%%%%%%%%%%%
%%%%%%%%%%%%%%%%%%%%%%%%%%%%%%%%%%%%%%%%%%%%%%%%%%%%%%%%%%%%%
%%%%%%%%%%%%%%%%%%%%%%%%%%%%%%%%%%%%%%%%%%%%%%%%%%%%%%%%%%%%%

\section{Conclusion}

In summary, we have presented a material-specific DFT\,+\,DMFT calculation of two intermetallic model materials, UGa$_2$ and UB$_2$, representing the extreme ends of the U\,5$f$ (de)localization scale. The parameters, Hubbard $U_{\rm ff}$, Hund's $J$ and, especially the double counting correction $\mu_{\rm dc}$, were tuned to best reproduce the valence band photoelectron spectroscopy data. Data from different incident photon energies enabled the disentanglement of subshells due to the different energy dependencies of the respective photoionization cross-sections. With the chosen parameters, the satellite structures of the U\,$4f$ core level spectra were also well reproduced, requiring only the adjustment of the core-valence interaction $U_{\rm fc}$. The DFT\,+\,DMFT calculation revealed multiple U\,5$f$ atomic configurations contributing to the ground states of UGa$_2$ and UB$_2$, with the average valence being nearly identical for both compounds. However, the width of the atomic configuration histograms and the time dependence of the charge correlation functions allowed for the assessment of itinerancy; the 5$f$ electrons in UB$_2$ were found to be much more itinerant than in UGa$_2$. This study paves the way to a systematic classification of uranium intermetallic compounds: it establishes an experiment-guided DFT\,+\,DMFT quantitative assessment of the charge state in this class of materials, as well as a quantitative simulation of the U\,$4f$ core-level spectra based on the Anderson impurity model coupled to the DFT\,+\,DMFT hybridization function. We argue that different spectroscopic techniques should yield consistent results for the charge state of uranium in intermetallics, and that the key ingredient is the proper inclusion of the electron correlation effects at the U sites relative to the itinerancy.

\section{Acknowledgements}

A.H.~was supported by JSPS KAKENHI Grant Numbers 21K13884, 21H01003, 23K03324, 23H03816, 23H03817, and the 2023 Osaka Metropolitan University (OMU) Strategic Research Promotion Project for Younger Researcher.
A part of the computational calculations were performed at the Vienna Scientific Cluster (VSC).
A.S. and M.M.F.C greatly acknowledge funding from the German Research Foundation (DFG) - grant N$^{\circ}$ 387555779.
D.T. acknowledges the support by the Deutsche Forschungsgemeinschaft (DFG) under the Walter Benjamin Programme, Projektnummer 521584902.
J.K. was supported by the project Quantum materials for applications in sustainable technologies (QM4ST), funded as Project No. CZ.02.01.01/00/22\_008/0004572 by Programme Johannes Amos Commenius, call Excellent Research.
We also acknowledge DESY — a member of the Helmholtz
Association HGF — for access to beamtime and support from the Max Planck-POSTECH- Hsinchu Center for Complex Phase Materials.

\section{Appendix}

\subsection{Finding $U_{\rm ff}$, $J$ and $\mu_{\rm dc}$ with valence band spectra}
\label{appendix_U_and_dc_scans}
\begin{figure*}[]
	\center
	\includegraphics[width=0.90 \linewidth]{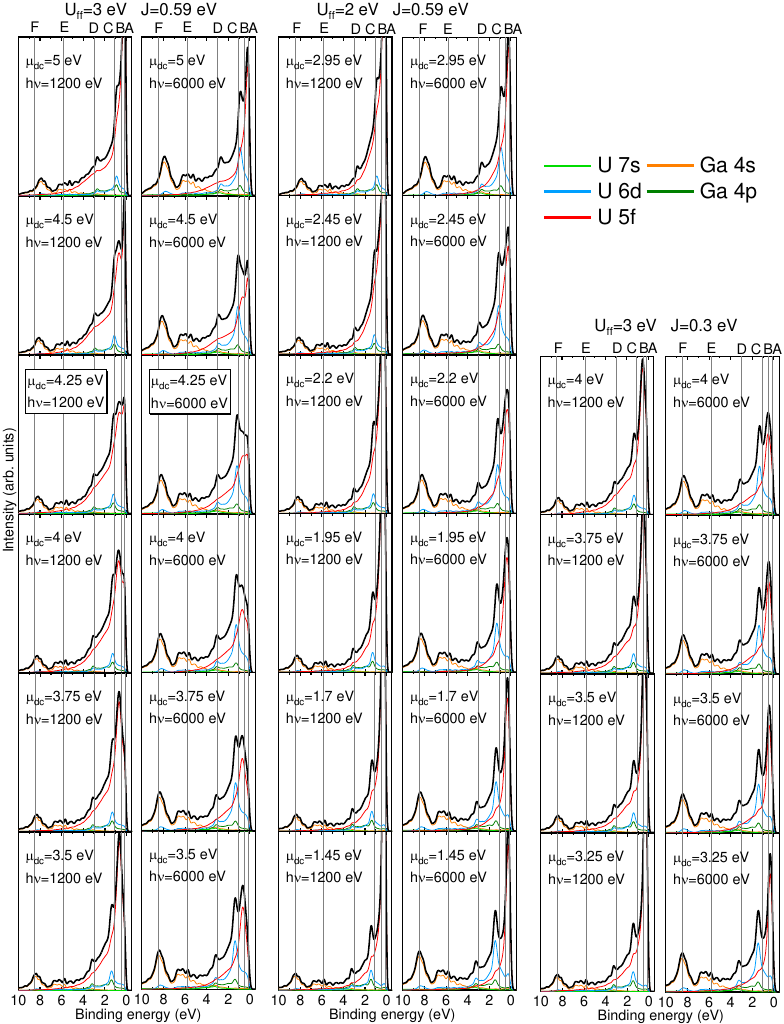}
	\caption{$U_{\rm ff}$ and $\mu_{\rm dc}$ dependence of the DFT\,+\,DMFT valence band spectra of UGa$_2$. The vertical lines denote the energy position of the experimental features $A$--$F$, see labeling at top of figure, and compare to Figure~\ref{fig_VB}~(a) and (b). The panels with the optimal value of the parameters fitting the experimental spectra are marked by a frame and show the same calculation as Figure~\ref{fig_VB}~(e) and (f).}
	\label{fig_dc_and_U_scans}
\end{figure*}

Figure~\ref{fig_dc_and_U_scans} summarizes the parameter dependence of the DFT\,+\,DMFT valence-band spectra for UGa$_2$. Analogous arguments can be made for UB$_2$. The gray vertical lines indicate the valence-band features ($A$ to $F$) in the experimental PES data, see Fig.~\ref{fig_VB}~(a) and (b). We used 2.0\,eV and 0.59\,eV as starting values for Hubbard $U_{\rm ff}$ and Hund's $J$, respectively, values that were used in the DFT\,+\,DMFT study of Chatterjee {\it et al}.\,\cite{Chatterjee21}. The $J$ value corresponds to 80\% of the Hartree-Fock value of the U\,5$f^3$ configuration\,\cite{Cowan1981}. Around these $U_{\rm ff}$ and $J$ values, we performed DMFT valence-band calculations for an extensive range of the double-counting correction $\mu_{\rm dc}$ values. 
For a given set of $U_{\rm ff}$ and $J$, the $\mu_{\rm dc}$ parameter controls the energy difference between U\,5$f$ and uncorrelated states. Thus, in the calculation, the energy positions that correspond to, e.g., the Ga\,4$s$ features $E$ and $F$ are almost linearly shifted to higher binding energies with increasing $\mu_{\rm dc}$. 

$\mu_{\rm dc}$, however, depends on the choice of $U_{\rm ff}$ and $J$ values because it is intended to subtract the electron-electron interaction effects present in the DFT result on the mean-field level. Thus, in Fig.~\ref{fig_dc_and_U_scans}, the $\mu_{\rm dc}$-dependent spectra for different sets of $U_{\rm ff}$ and $J$ are arranged in a matrix manner, such that in each row, the calculated $F$ feature appears at a similar binding energy. This demonstrates that one can find values for $\mu_{\rm dc}$ that reproduce the high-energy valence-band features for any set of $U_{\rm ff}$ and $J$. However, the choice of the $U_{\rm ff}$ and $J$ affects the U $5f$ features through the $f$--$f$ interaction. 
It is important to note for $U_{\rm ff}=3.0$~eV and $J=0.59$~eV additional good agreement with the experimental U 5$f$ features $A$ and $B$ is obtained.

\begin{figure*}[t!]
	\center
	\includegraphics[width=0.8\linewidth]{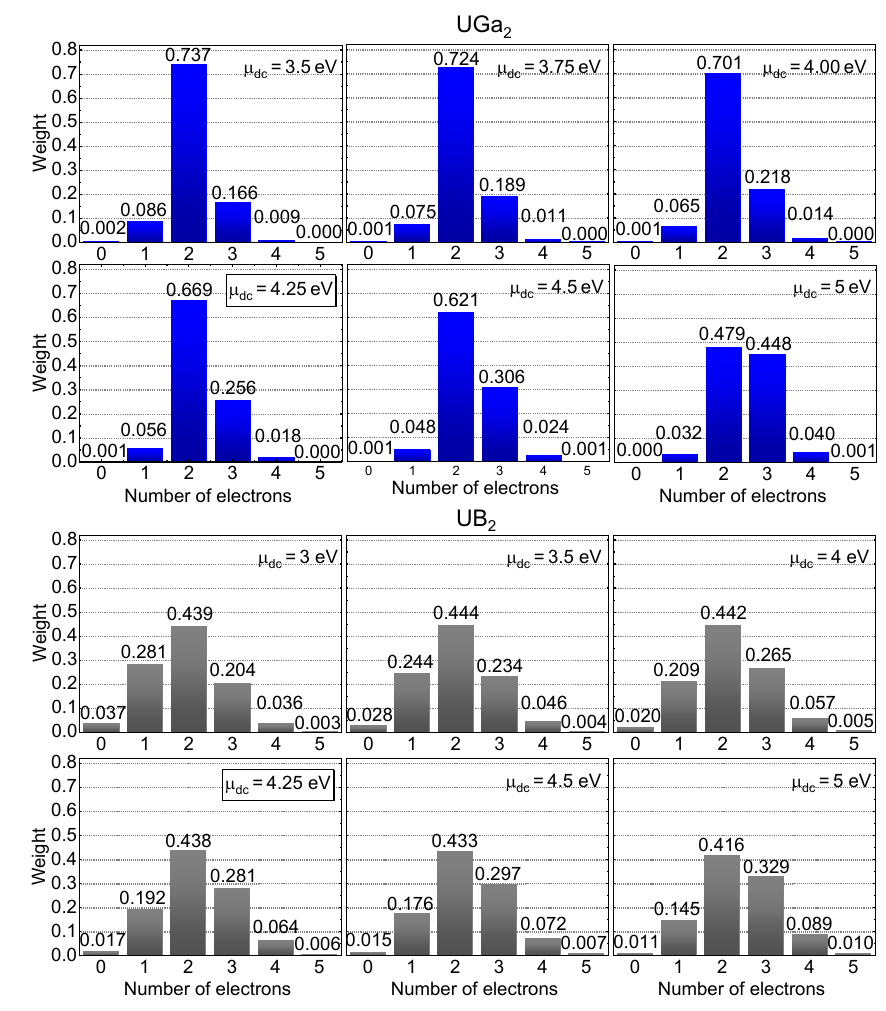}
	\caption{  $\mu_{\rm dc}$ dependence of the atomic configuration histograms of UGa$_2$ (top) and UB$_2$ (bottom) , where $U_{\rm ff}=3$~eV and $J=0.59$~eV. The panels with the optimal value of the paramters fitting the experimental spectra are marked by a frame and show the same data as Figure~\ref{Fig_histograms_and_charge_corr}~(a) and (b).}
	\label{Fig_histograms_all}
\end{figure*}

\begin{figure*}[t!]
	\includegraphics[width=1\linewidth]{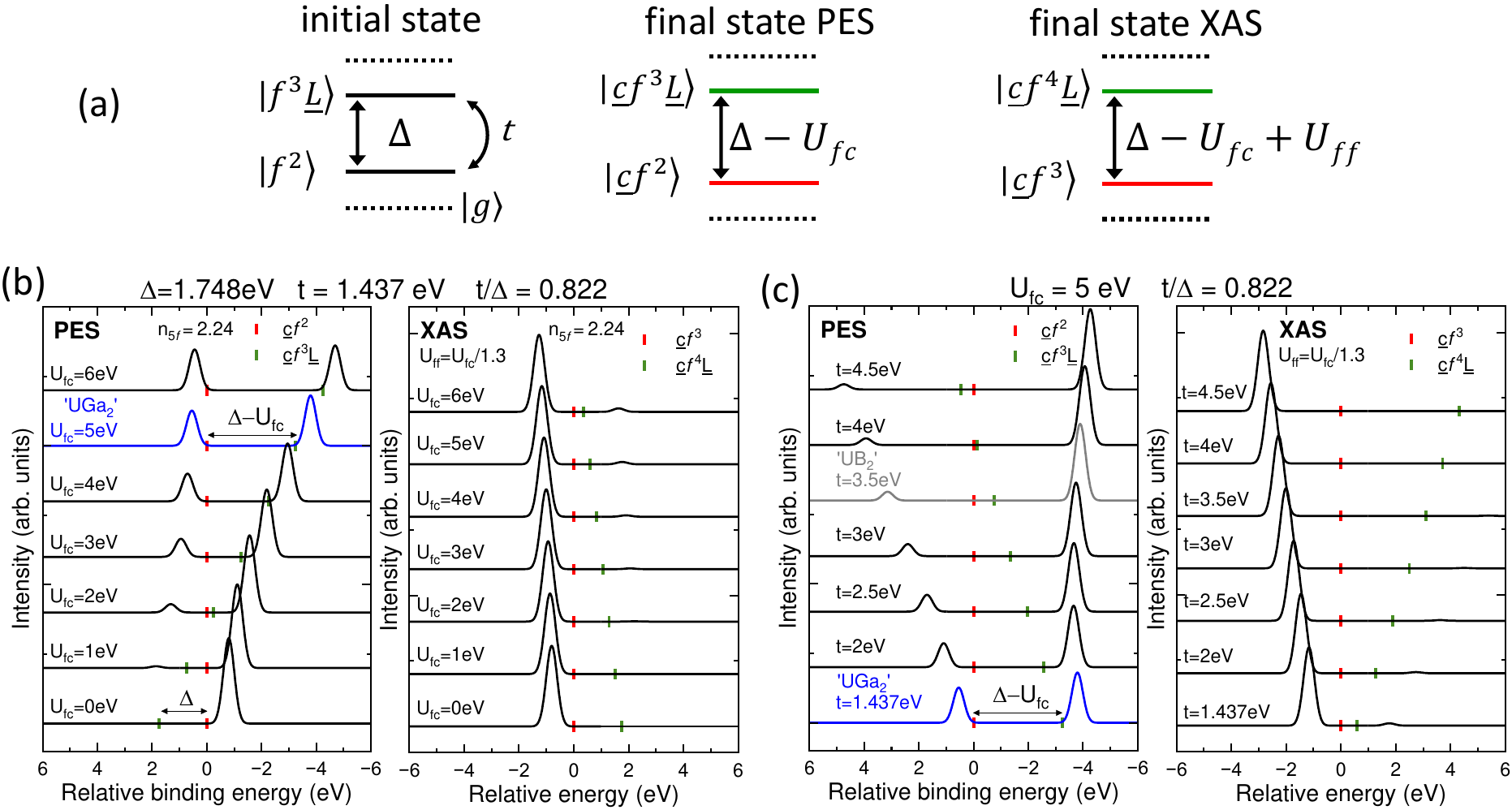}
	\caption{(a) Schematic energy diagram displaying the initial state problem with ground state $|g\rangle = \alpha |f^2\rangle + \beta |f^3\underline{L}\rangle$, and PES and XAS final states. (b) Core-level (\underline{c}) PES and XAS spectra in the two-configuration toy-model, maintaining the same average 5$f$ filling, for various core-valence interaction values $U_{\rm fc}$ and a fixed small hopping $t$ (and small $\Delta$) addressing the UGa$_2$ electronic structure problem (see blue spectrum). In (c) core-level PES and XAS spectra in the two-configuration toy-model, maintaining the same average 5$f$ filling, for several values of hopping $t$ (and $\Delta$) with fixed $U_{fc}$\,=\,5\,eV, addressing the problem of UB$_2$ (see gray spectrum). The red and green ticks mark the energy positions of the non-hybridized final states. }
			\label{Fig_toy_diagram}
\end{figure*}

\subsection{$\mu_{\rm dc}$ dependence of the atomic configuration histograms}
\label{appendix_histograms}

Figure~\ref{Fig_histograms_all} shows the variation of the atomic configuration histograms of UGa$_2$ and UB$_2$ with $\mu_{\rm dc}$. The double-counting correction renormalizes the energy position of the U\,5$f$ states, thus having a large impact on the valence distribution. This is particularly evident for UGa$_2$, where the distribution is very narrow for a shallower U\,$5f$, e.g., for $\mu_{\rm dc}$\,=\,3.50~eV, the 5$f^2$ configuration strongly dominates (73.7\%), while for deeper 5$f$ states, e.g., for $\mu_{\rm dc}$\,=\,5~eV, the $f^2$ and $f^3$ configurations contribute almost equally to the ground state.

\subsection{Toy-model for comparing core-level PES and XAS}
\label{appendix_toymodel}

%%%%%%%%%%%%%%%%%%%%%%%%%%%%%%%%%%%%%%%%%%%%%%%%%%%%%%%%%%%

%%%%%%%%%%%%%%%%%%%%%%%%%%%%%%%%%%%%%%%%%%%%%%%%%%%%%%%%%%% 
We consider a two-level \textit{initial state} Hamiltonian $\hat{H}$ consisting of $f^2$ and $|f^3\underline{L}\rangle$ configurations, with $\underline{L}$ representing a valence hole. The two levels are separated by the charge transfer energy $\Delta$ and are coupled by the hopping amplitude $t$ (see Initial Fig.\,\ref{Fig_toy_diagram}\,(a))\,\cite{gunnarsson1983,degroot1994,Kotani1988}. In the toy model
%%%%%%%%%%%%%%%
\begin{equation*}
	\hat{H} =
	\begin{pmatrix}
		0 & t \\
		t & \Delta   
	\end{pmatrix}
\end{equation*}
%%%%%%%%%%%%%%%%
the ground state $|g\rangle = \alpha |f^2\rangle + \beta |f^3\underline{L}\rangle$ with coefficients $\alpha$ and $\beta$ is determined by ratio $t/\Delta$. We select $t/\Delta$\,=\,0.822, resulting in the average 5$f$ occupation $n_{5f}=2.24$ for UGa$_2$ and UB$_2$ that was obtained in the main text, with $\alpha^2$ and $\beta^2$ amounting to 0.76 and 0.24, respectively.

In the following two cases will be discussed: 
1) How $U_{\rm fc}$, for a fixed choice of $t$ and $\Delta$, and fixed ratio of $U_{\rm ff}$/$U_{\rm fc}$ in case of XAS, impacts the satellites in core-level PES and XAS. 2) How $t$ and $\Delta$, for the same ratio $t/\Delta$\,=\,0.822 and a given set of $U_{\rm ff}$ and $U_{\rm fc}$, impact the core level satellite in PES and XAS.

%%%%%%%%%%%%%%%%%%%%%%%%%%%%%%Case 1%%%%%%%%%%%%%%%%%%%%%%%%%%%%%%%%%%%
\underline{Case 1)}: The final states of the core-level PES and XAS processes are characterized by the presence of a core hole, necessitating consideration of the  core-valence interaction $U_{\rm fc}$ between a core hole ($\underline{c}$) and 5$f$ valence electrons. 

In the PES \textit{final state} charge neutrality is broken. The basis is formed by $|\underline{c}f^2\rangle$ and $|\underline{c}f^3\underline{L}\rangle$ and the Hamiltonian $\hat{H}_{\rm PES}$ reads: 
%%%%%%%%%%%%%%%%
\begin{equation*}
	\hat{H}_{\rm PES} =
	\begin{pmatrix}
		0 & t \\
		t & \Delta-U_{\rm fc}
	\end{pmatrix}.
\end{equation*}
%%%%%%%%%%%%%%%%
In the final state, the configurations $|\underline{c}f^3\underline{L}\rangle$ and $|\underline{c}f^2\rangle$ are separated by $\Delta - U_{\rm fc}$ (see \textit{final state PES} in Fig.~\ref{Fig_toy_diagram}\,(a)) and red and green ticks in the left panel of Fig.~\ref{Fig_toy_diagram}\,(b). Choosing $\Delta$\,=\,1.75~eV with $t$\,=\,1.4373~eV, maintaining the ratio $t/\Delta$\,=\,0.822, we calculate the core-level spectra for several values of $U_{\rm fc}$ between 0 and 6\,eV (see left panel of Fig.~\ref{Fig_toy_diagram}\,(b)). The exercise shows the sensitivity of peak positions and intensity ratios to $U_{\rm fc}$ for a given $t$ and $\Delta$. For $U_{\rm fc}$\,=\,0, the energy separation of the red and green ticks is given by $\Delta$. With increasing $U_{\rm fc}$, the levels eventually cross when $\Delta-U_{\rm fc}$ becomes negative, i.e., the order of configurations is reversed with respect to the initial state. For $U_{\rm fc}$\,=\,5\,eV (see blue curve in Fig.~\ref{Fig_toy_diagram}\,(b)), the value used in the present DFT\,+\,DMFT calculations, the toy model mimics the strong satellite of the 4$f_{7/2}$ PES spectrum of UGa$_2$ in Fig.~\ref{Fig_core_level}. With $U_{\rm fc}$ larger than $\Delta$, the order of states is thus reversed in UGa$_2$. We note, that this is actually a typical situation for the core level spectra of $f$ electron materials. 

The relative spectral weights in the final state are not equal to the relative abundances $\beta^2$ and $\alpha^2$ in the initial state. The transfer of spectral-weight originates from the superposition (quantum-mechanical interference) of the $f^2$- and $f^3$-derived states in both the initial and final states. Therefore, the PES spectra cannot be modeled with an $f^2$ ionic model, even though the $f^2$ dominates the initial state.  The reversed order of the configurations in the final state leads to substantial redistribution of spectral intensities. Only in the limit of very large $U_{\rm fc}$ or very small $t$ would the  PES intensities correspond to the  weights of the initial state configurations. However, for uranium intermetallics both covalence and Coulomb interaction must be treated on the same footing.

For the charge neutral process in XAS, the basis is formed by $|\underline{c}f^3\rangle$ and $|\underline{c}f^4\underline{L}\rangle$, and the Hamiltonian $\hat{H}_{\rm XAS}$ reads:
%%%%%%%%%%%%%%%%
\begin{equation*}
	\hat{H}_{\rm XAS} =
	\begin{pmatrix}
		0 & t \\
		t & \Delta-U_{\rm fc}+U_{\rm ff}
	\end{pmatrix}.
\end{equation*}
%%%%%%%%%%%%%%%%
For the $|\underline{c}f^4\underline{L}\rangle$ configuration, the energy gain due to $U_{\rm fc}$ is compensated by the extra $f$--$f$ Coulomb repulsion $U_{\rm ff}$ due to the extra (photo-excited) $f$ electron (see \textit{final state XAS} in  Fig.~\ref{Fig_toy_diagram}\,(a)). Typically, $U_{\rm fc}$ is larger than $U_{\rm ff}$, but they are of the same order of magnitude. Thus, in XAS, the levels in the final state have the same order as in the initial state, unlike in PES (see red and green ticks in right panel of Fig.~\ref{Fig_toy_diagram}\,(b)). 

We fix the parameter $U_{\rm ff}$ to $U_{\rm fc}$/1.3 and then calculate the XAS spectra for several values of $U_{\rm fc}$ between 0 and 6\,eV (see Fig.~\ref{Fig_toy_diagram}~(b)). The main line and satellite are composed largely of $|\underline{c}f^3\rangle$ and $|\underline{c}f^4\underline{L}\rangle$, respectively, with the main line consistently dominating the spectrum, despite the mixture of $f^2$ and $f^3$ in the initial state. Hence, XAS spectra reflect the dominant $f^2$ configuration. In contrast to PES, the satellites are never very large and peak positions and intensities are fairly insensitive to $U_{\rm fc}$. In case $U_{\rm ff}$\,=\,$U_{\rm fc}$, the satellite intensity would even be zero for all values of $U_{\rm fc}$

Therefore, an ionic model considering only the formal $f^2$ configuration provides a valid description for core-level XAS (similarly for NIXS) in uranium intermetallic compounds.

%%%%%%%%%%%%%%%%%%%%%%%%%%%%%%Case 2%%%%%%%%%%%%%%%%%%%%%%%%%%%%%%%%%%%
\underline{Case 2)}: Now, we discuss the impact of $t$ and $\Delta$ with a fixed $t/\Delta$ of 0.822 to maintain the same 5$f$ occupation. We now fix $U_{\rm fc}$\,=\,5\,eV and vary $t$ (and $\Delta$) from 1.437\,eV, the value mimicking the case of UGa$_2$ (see blue curve in Fig.~\ref{Fig_toy_diagram}~(c)), to $t$\,=\,4.5\,eV. Again, we observe a strong dependence of position and intensity ratios on the actual size of $t$ (and $\Delta$).  The hopping of $t$\,=\,3.5\,eV mimics the PES core-level data of UB$_2$ (see gray curve in Fig.~\ref{Fig_toy_diagram}~(c)). For even larger values of $t$ (and $\Delta$), the satellite becomes even smaller and the order of states reverses back to the same order as in the initial state. In XAS, as before, the spectra exhibit only one major line and a negligible satellite spectral weight, the latter becoming even less pronounced the larger $\Delta$ becomes with respect to $U_{\rm fc}$ and $U_{\rm ff}$.

\clearpage

%\bibliography{references_a}

\end{document}